RESEARCH ARTICLE

# Navigation Strategies of Motor Proteins on Decorated Tracks

Zsolt Bertalan[1], Zoe Budrikis[1], Caterina A. M. La Porta[2]*, Stefano Zapperi[1,3,4,5]*

**1** Institute for Scientific Interchange Foundation, Torino, Italy, **2** Center for Complexity and Biosystems, Department of Bioscience, University of Milan, Milano, Italy, **3** Center for Complexity and Biosystems, Department of Physics, University of Milan, Milano, Italy, **4** CNR - Consiglio Nazionale delle Ricerche, Istituto per l'Energetica e le Interfasi, Milano, Italy, **5** Department of Applied Physics, Aalto University, Aalto, Espoo, Finland

\* caterina.laporta@unimi.it (CAMLP); stefano.zapperi@unimi.it (SZ)

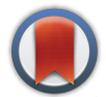

**Data Availability Statement:** All relevant data are within the paper and its Supporting Information files.

**Funding:** Zsolt B., Zoe B., and SZ are supported by the ERC advanced grant SIZEFFECTS. SZ acknowledges partial support from the Academy of Finland FiDiPro progam, project 13282993. The funders had no role in study design, data collection and analysis, decision to publish, or preparation of the manuscript.

**Competing Interests:** The authors have declared that no competing interests exist.

## Abstract

Motor proteins display widely different stepping patterns as they move on microtubule tracks, from the deterministic linear or helical motion performed by the protein kinesin to the uncoordinated random steps made by dynein. How these different strategies produce an efficient navigation system needed to ensure correct cellular functioning is still unclear. Here, we show by numerical simulations that deterministic and random motor steps yield different outcomes when random obstacles decorate the microtubule tracks: kinesin moves faster on clean tracks but its motion is strongly hindered on decorated tracks, while dynein is slower on clean tracks but more efficient in avoiding obstacles. Further simulations indicate that dynein's advantage on decorated tracks is due to its ability to step backwards. Our results explain how different navigation strategies are employed by the cell to optimize motor driven cargo transport.

## Introduction

Many essential cellular processes rely on active transport. Examples include spindle positioning [1, 2] and transport of cargoes like chromosomes [3–6] and organelles [7–9]. These tasks are performed by motor proteins, which are molecules that convert chemical energy into mechanical energy that is used to travel along molecular tracks such as microtubules. Different motor proteins display distinct patterns of motion, the details of which have only been elucidated in recent years with the development of novel quantum dot-based experimental techniques for tracking molecules [10].

For example, kinesins are microtubule-based motors that move in a coordinated and precise manner [9]. That motion can be linear, such as in the case of wild-type kinesin-1 [11], or helical with fixed chirality for other kinesin types and specially engineered kinesin-1 [12, 13]. In contrast to the regular motion of kinesin, the motor protein dynein [14] takes uncoordinated random steps [15, 16], and moves helically but with random changes of chirality [17]. A recent review of models for active transport in the cell can be found in Ref. [18].

These different approaches to motion on a track should offer various advantages and disadvantages. *In vitro* studies show that kinesins proceed faster along the microtubule than dynein, $\sim$ 400 nm/s for kinesin-1 [19], compared to $\sim$ 120 nm/s for dynein [15, 20]. *In vivo*, however, microtubules are rarely clean and are instead decorated with microtubule-associated proteins, other motor proteins and cofactors. These objects can act as roadblocks during the transport of intracellular cargoes and thus inhibit motor motility [21, 22]. Indeed, kinesin-1 is effectively inhibited by any kind of obstacle [22], whereas it has been speculated that helically moving kinesin [13] and dynein [17] should be able to navigate the microtubule by off-axis stepping and/or backwards stepping, without the need of detachment/attachment events. This should be of advantage for dynein, which is measured as having lower probability of successful multiple re-attachment events than the kinesins [20]. However, these speculations have not been subject to comparative quantitative testing.

Here, we present a model for motor protein motion on a track, incorporating the most fundamental observations about the stepping patterns of the various motor proteins. By simulating motor protein motion on tracks on which random areas are inaccessible we clarify the role of stochastic stepping in navigation of obstacles. Our simulations confirm that both off-axis and backwards steps are needed for successful navigation of crowded tracks, but that the backwards motion of dynein is a particularly important part of its strategy.

## Methods

We model motor proteins as two heads that walk on a microtubule lattice with 13 columns that represent protofilaments, with periodic boundary conditions. Each protofilament has 1000 sites along it, representing the 8nm long tubulin subunits. At each time interval $dt$ we determine whether the motor will step with the probability $r_{step} dt$ and then perform a step according to rules for each motor protein, described below. Our model for motor protein stepping is illustrated in Fig 1. Parameters for dynein and kinesin are reported in Table 1. Parameters are drawn from experimental observation, with the exception of $r_{step}$ which is extrapolated from data for lower ATP concentrations, as described below.

We measure velocities by preparing a track and a motor protein to walk on it. The position of the motor is determined by an average of both heads. The velocity is determined every 100 Monte Carlo steps from the position of the motor protein, until either the motor gets stuck (no valid moves for either head), it reaches the end of the track, or the end of the simulation time is reached. Averages are made over 1000 simulation runs. Movies of typical trajectories for dynein (S1 Video) and helical kinesin (S2 Video) on clean tracks and dynein (S3 Video) and helical kinesin (S4 Video) on decorated tracks are given as Supporting Information.

### Kinesin stepping

There are many types of kinesin, but we focus on those whose steps are deterministic. The motor steps towards the plus-end of the microtubule, in a manner not unlike bi-pedal walking, with the two heads stepping alternately. The distribution of step sizes is narrow and with a peak at 16 nm [10, 24], resulting in 8nm advances of the motor center-of-mass per step. The rate of stepping depends on the concentration $c$ of ATP. For low concentrations, $c \ll 150$ μM, the velocity is approximately proportional to the concentration [25, 26], which leads to a stepping rate of $r_{step} \approx 3.3$/s at level of 1 μM ATP. We chose this concentration since it is representative for many experiments and to be able to compare to the properties of dynein at similar levels of ATP.

For some types of kinesin, such as kinesin-2 and kinesin-8, a small fraction $0.05 \lesssim p_{ang} \lesssim$ 0.3 of steps have a fixed sideways, or "off-axis" component [12, 13, 19, 23], with the value of

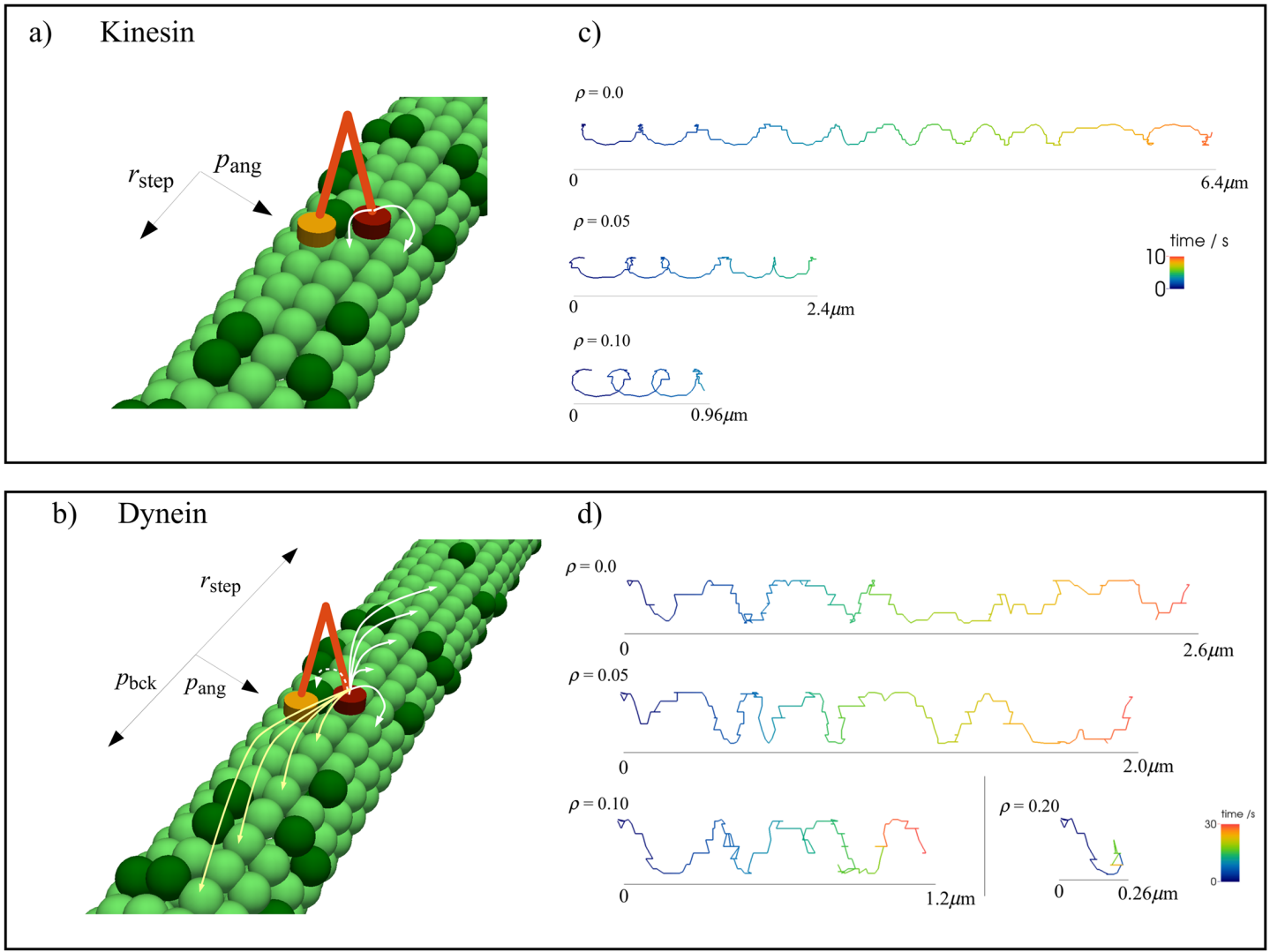

**Fig 1. Model for motor protein stepping on a microtubule track.** The track is represented as a square lattice of microtubule subunits on which the heads of the motor proteins process. Dark green sites represent subunits that are inaccessible to the motor protein. The stepping per head occurs with rate $r_{step}$. A: Kinesin heads alternate in taking steps of 8 nm in the forward direction (towards the plus end of the microtubule). B: Each dynein head has an equal chance to step, allowing one head to take multiple consecutive steps. Dynein can take forward steps (towards the minus end of the microtubule) and, with probability $p_{bck}$, backward steps (toward the plus end). Dynein steps have size up to $4 \times 16$ nm. Both motors take sideways steps with probability $p_{ang}$, changing only one protofilament, but while the off-axis direction of kinesin is fixed, the direction of sideways steps of dynein is determined by the leading head, indicated by the dashed arrows. C: Sample trajectories of kinesin for different obstacle concentrations $\rho$. D: Sample trajectory of dynein for different $\rho$.

doi:10.1371/journal.pone.0136945.g001

Table 1. Summary of the parameters employed in the model.

| Name | Symbol | Kinesin | Dynein | Comment |
|---|---|---|---|---|
| stepping rate | $r_{step}$ | 3.3/s | 0.14/s | 1 µM ATP |
| off-axis prob. | $p_{ang}$ | 0.05-0.3 | 0.2-0.4 | [12, 23], [16] |
| backwards-step | $p_{bck}$ | - | 0.2 | [15] |
| stepping head | - | alt. | 1/2 | [10], fits Ref. [16] |
| stepping size | s | 8nm | 16$n$ nm, $n$ = 1,2,3,4 | [10], [16] |

doi:10.1371/journal.pone.0136945.t001

$p_{\text{ang}}$ set by the neck-linker length [12, 27]. In this case, motion around the microtubule is helical with pitch approximately 500-1000nm [12]. This helical motion is very regular as well and has a conserved chirality [12, 13]. In most of our simulations, we used $p_{\text{ang}} = 0.2$, but we have also tested a range of $p_{\text{ang}}$ values. (In what follows, "kinesin" refers to these types of kinesin unless otherwise specified.)

By way of contrast, wild-type kinesin-1 moves strictly along a straight trajectory [11], although very recent experiments indicate it can shift sideways by rapid unbinding/binding events [28]. We have also simulated a motor protein confined to a single track in order to determine the role of chiral motion in kinesin-2 and kinesin-8.

## Dynein stepping

Experiments show that dynein proceeds towards the minus-end of the microtubule—the opposite direction to kinesins—but dynein heads also step backwards, i.e. towards the plus-end, with an experimentally observed probability of $p_{\text{bck}} = 0.2$ [15, 16, 29]. Furthermore, dynein steps are not hand-over-hand, and do not alternate regularly. Instead, dynein proceeds in a shuffling manner, where a lagging head trails a leading one, with the change of role infrequent, and it is possible for one head to take two (or more) consecutive steps [16]. These steps are taken with a rate of $r_{\text{step}} = 0.14/\text{s}$ [15, 30] at $1\mu M$ ATP, and the stepping head is determined randomly with a probability of 0.5 for each head, yielding the observed consecutive-step probability of 0.25 in line with experimental observations [16].

In the model, the dynein-step size $s$ on a clean microtubule is randomly distributed as $s = 16n$ nm, where $n$ is exponentially distributed and $1 \leq n \leq 4$. This is derived from the observation that dynein steps correspond to the consumption of a molecule of ATP [31] and we hypothesize that the step size is proportional to the number of ATP molecules bound within the time $dt$, motivated by the observation that step size distributions are dependent on the ATP concentration [29, 32]. Dynein can bind up to four ATP molecules [32] and for fixed ATP adsorption rate $r_{\text{ads}}$ the probability of $n$ molecules adsorbed in time $dt$ is exponentially distributed

$$P(n) \sim (r_{\text{ads}}dt)^n \tag{1}$$

Off-axis steps are taken with probability $0.2 \leq p_{\text{ang}} \leq 0.4$ in the direction of the leading head [16]. This leads to helical motion with a pitch of $\approx 500$ nm [17], but, in contrast to kinesin, the chirality of the trajectories is not conserved [17]. Therefore, we hypothesize that the off-axis motion for dynein tends to be in the direction of the leading head due to chemical or structural reasons, in other words, if the left head is leading, the sideways component of the step will be to the left. This is unlike the off-axis component of kinesin steps, which are due to structural asymmetries in the neck-linker domains [12, 23] and are always in the same direction.

## Restrictions on motion and decorated tracks

Both motor proteins are also subject to the following restrictions on their motion: A step is not allowed if the target site is occupied by the not-stepping head. Instead, the closest available site to the previous target in direction of the step direction is chosen. Also the two heads are not allowed to be separated by distances larger than a few tens of nm. To be precise, the maximum head-to-head distance on the microtubule lattice for kinesin is chosen to be 24nm, while for dynein the maximum distance is 72nm. These values are chosen according to be slightly larger than the maximum modelled step sizes.

Decorated microtubules have sites made inaccessible with a uniform probability per site given by the decoration fraction $\rho$. At each time step, after determining the stepping head and

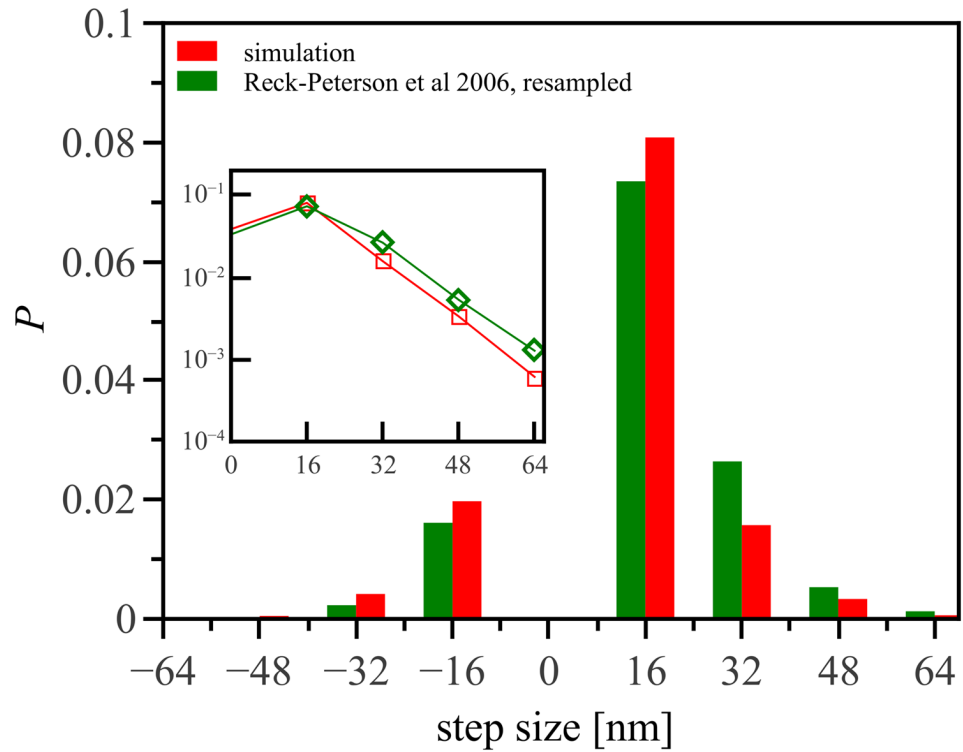

**Fig 2. Comparison of step size distributions in simulations with experimental results from Ref. [15].**
The global maximum of both distributions is at 16nm, and a significant portion (0.2) of the steps are in the
backward direction. In the experiment step sizes were sampled in small bins of width 2 nm, while the
simulation allows only steps of size 16 nm only for dynein. We therefore re-binned the (digitized) experimental
data for comparability purposes, shown in green. Inset: the distributions on a log scale. In both experiments
and simulation, the tail of the distribution is experimental. For clarity, only the positive step sizes are shown,
but negative steps are also exponentially distributed.

direction and step size of the motor, we check whether the path between the head's initial and
final sites contains inaccessible sites. If it does, the stepping head first makes as much of its on-
axis step as possible. If the step has an off-axis component, the sideways step is then taken, if
possible. Finally, if the original step involved more on-axis motion, the rest of that step is exe-
cuted, if possible. We note in particular that under this scheme, dynein can take steps that are a
multiple of 8 nm (the track lattice constant) on decorated tracks.

## Results

### Dynein step size distribution

We first confirm that our model reproduces observed stepping behavior of dynein, without fit-
ting parameters. Fig 2 shows step size distributions for experiments (obtained from Ref. [15])
and our simulations. These simulations are performed on a clean 13-protofilament microtu-
bule. Under these conditions our stepping rules for dynein allow only steps in multiples of
16nm. To be able to compare the experimental result to our simulation, we re-binned the
experimental data and see that the simulations reproduce very well the observed step-size dis-
tributions. The inset of Fig 2 shows the distribution of positive step sizes on a logarithmic scale,
and confirms that the distribution tails fall off exponentially for both simulation and experiments. This is consistent with our stepping model in which the step size is proportional to the exponentially distributed number of ATP molecules adsorbed in a time step (Eq (1)).

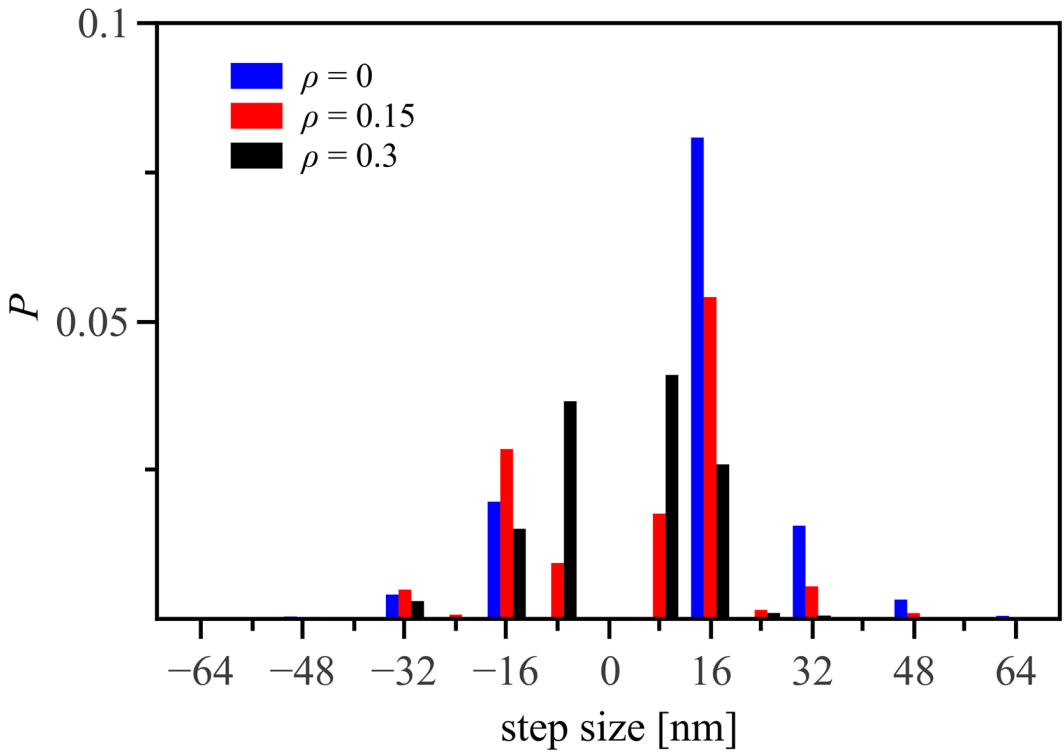

**Fig 3. Step-size distribution of dynein on decorated tracks.** While the decoration fraction $\rho = 0.15$ (red) shows qualitatively the same results as the clean case, $\rho = 0$ (blue), a decoration fraction of $\rho = 0.3$ (black) differs significantly in shape from the previous curves. The global maximum of the distribution shifts from 16nm to 8nm and there are approximately as many steps taken forward as backward, showing that the motor stays mainly in a single area and does not progress along its track.

## Dynein step size distribution on decorated tracks

Because dynein can move in all directions, we expect that it is able to also navigate tracks with obstacles. We have simulated motion of dynein on a 13-protofilament microtubule on which a fraction of the possible binding sites for heads are made inaccessible. The distribution of step sizes yields information about both local and global motion of the motor protein. As shown in Fig 3, for small decoration fractions $\rho < 0.1$ the distribution is strongly peaked at +16 nm, in other words most steps are taken forwards. However, as $\rho$ increases, the majority step size shifts to 8 nm, which corresponds to a single tubulin block. Such a step is only taken in our simulations if a multiple of 16 nm (two tubulin blocks) is not possible. Furthermore, the step size distribution becomes increasingly symmetric, that is, the weight of the distribution in positive and negative directions becomes approximately equal. This indicates that although the dynein motor continues to move locally, its global progress is hampered by obstacles.

We find that dynein motion is ultimately blocked completely for $\rho \gtrsim 0.3$. Analysis of the distribution of clusters of inaccessible sites indicates this corresponds to a decoration fraction where around 1% of clusters have width $w = 13$—that is, the motor has substantial probability to encounter clusters whose size perpendicular to the track axis spans the whole track. In other words, dynein's navigation strategy finds a path forwards as long as such a viable path exists.

## Velocity of dynein and kinesin on decorated tracks

We next examine the differences between dynein and kinesin, in terms of how their stepping behaviour affects navigation of obstacles and velocity of the motors. Unlike dynein which steps stochastically, when kinesin encounters an obstacle ahead of it, it takes a purely sideways step in the direction determined by the fixed chirality of its motion. If this is not possible the motor comes to a halt, is considered to be stuck, and in course detaches from the track. On the other hand, on clean tracks kinesin always steps forward and its deterministic motion should be faster than the stochastic stepping of dynein.

Fig 4 shows the mean velocity $\langle v \rangle$ plotted against microtubule decoration fraction for kinesin and dynein. On a clean track, kinesin is significantly faster, with dynein velocity of 800 nm/s compared to 420 nm/s for dynein, values which agree with experiments [15, 19, 30]. On decorated tracks, however, the velocity of kinesin decreases sharply with decoration fraction, but the slowdown is much less pronounced for dynein. For decoration fractions above $\rho \sim 0.025$, wild-type dynein is faster than wild-type helical kinesin, indicating that backwards stepping motion does confer an advantage on decorated tracks.

Beyond the mean velocity, our simulations also yield information about the distribution of instantaneous velocities $v$, whose distributions are reported in Fig 5. At zero decoration fraction, the velocities of dynein and kinesin have a normal distribution about their mean value. However, on decorated tracks the distributions become bimodal and develop peaks around $v = 0$, which increase in weight as the decoration fraction increases, reflecting the increasing difficulty of navigation.

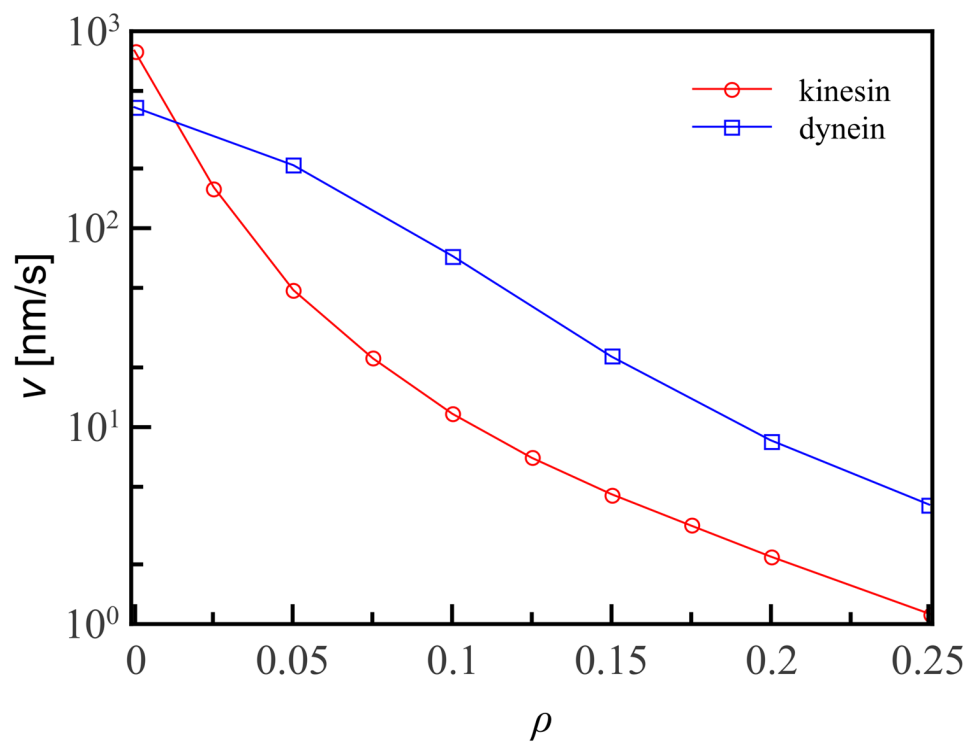

**Fig 4. Average velocity of motor proteins *vs* decoration fraction $\rho$.** The velocity of kinesin decreases by an order of magnitude as $\rho$ increases from $\rho = 0$ to $\rho = 0.05$, but the velocity of dynein decreases only slowly with the decoration fraction. Plotted here are velocities for wild-type motor proteins, with $p_{ang} = 0.4$. Lines are guides to the eye; errorbars are smaller than the symbol size.

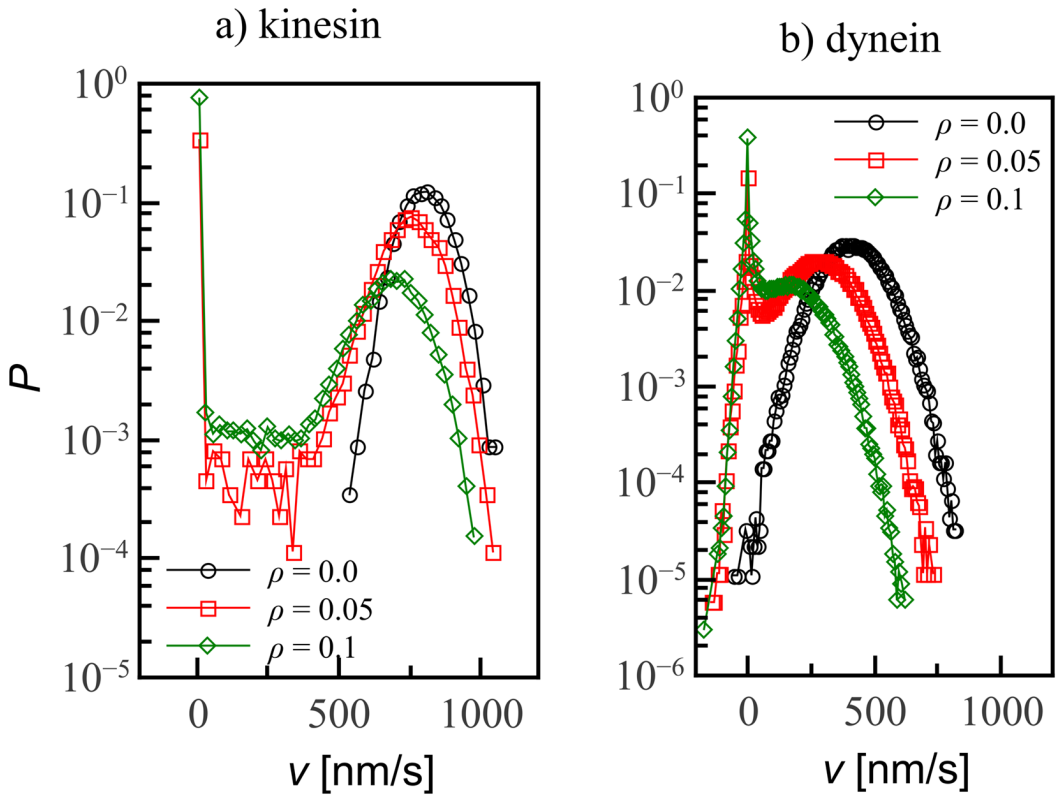

**Fig 5. Velocity distributions are normal on undecorated tracks but develop a peak around $v = 0$ as decoration fraction $\rho$ is increased.** A: Kinesin, whose velocity can only be positive. B: For dynein, the distribution includes also negative velocities. Lines are guides to the eye; errorbars are smaller than the symbol size.

## Role of off-axis and backward steps in stochastic motion

Even for the coordinated stepping of wild-type kinesin, helical motion confers and advantage over linear motion. Simulations of kinesin-1, which moves along a single protofilament, indicate that its motion is halted for any decoration fraction. The importance of sideways steps for navigation has also been identified experimentally [28]. We have tested a range of values for the off-axis parameter $0.02 \leq p_{ang} \leq 0.6$ away from its wild-type values. We find that for $p_{ang} \gtrsim 0.1$ the fraction of off-axis steps does not make a large contribution to the motor protein velocity on decorated tracks, for both motor proteins, as shown in Fig 6.

As well as being able to take off-axis steps in both directions, dynein is different from kinesin in that it can step backwards. To illuminate the role of these backwards steps, we have simulated a modified kinesin model in which the walker can take a small fraction of its steps backwards. As shown in Fig 7, this increases its abilities to navigate a decorated track substantially. Indeed, for large $\rho > 0.15$, backward-stepping kinesin moves faster even than wild-type dynein. This suggests that the relevant aspect of dynein's advantage over wild-type kinesin at large decoration fraction is its backwards stepping, rather than stochasticity in the off-axis steps.

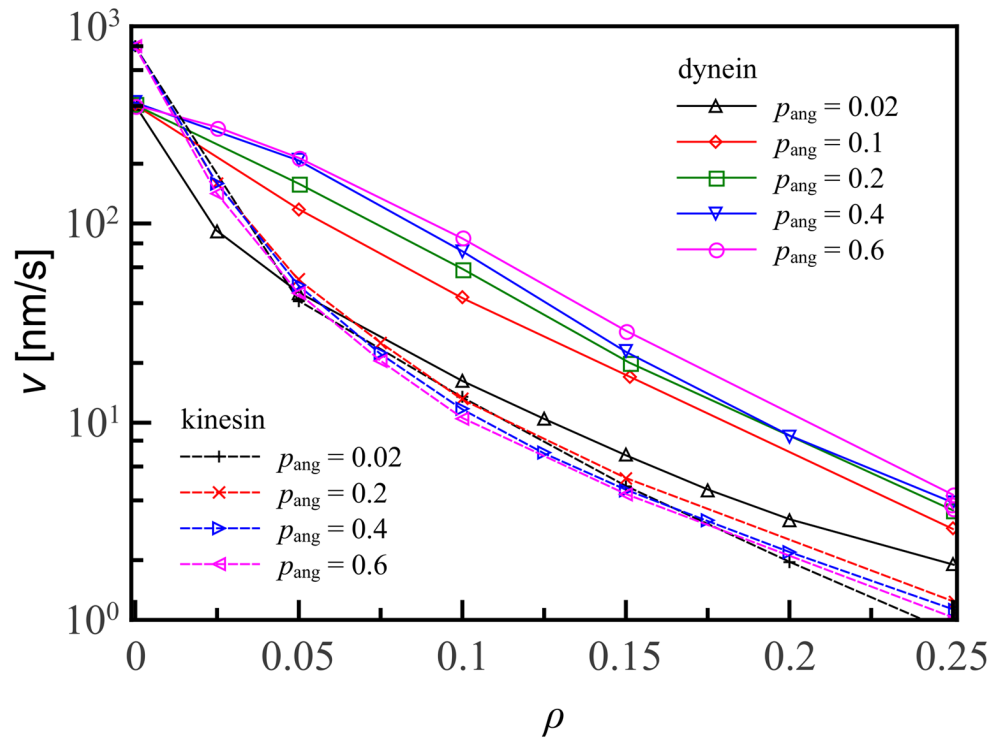

**Fig 6. Velocity profiles for kinesin and dynein are largely unaffected by changes in the off-axis parameter $p_{ang}$.** For the range of $p_{ang}$ studied, $0.02 \leq p_{ang} \leq 0.6$, changing $p_{ang}$ does not affect the shape of the velocity profile except in the case of dynein with $p_{ang}$ = 0.02, where dynein is faster than kinesin only for large $\rho$. Lines are guides to the eye; errorbars are smaller than the symbol size.

## Discussion

Although the stepping properties of motor proteins have been subject to intense experimental scrutiny in recent years [10, 12, 13, 15–17, 19, 27–30], until now there has been lacking a theoretical model that unifies diverse experimental observations. We have implemented a motor stepping model based on insights gained from experiments, with only the stepping rate as a fitting parameter. Our simulations show that while the stochastic motion of dynein makes it slower than helically-moving kinesin on clean tracks, as the density of roadblocks is increased dynein continues to navigate successfully, whereas kinesin becomes trapped. We are also able to quantitatively compare how the differences in stepping patterns of different motor proteins affect their ability to navigate obstacles and to identify key features of their motion that enable navigation.

These findings provide a perspective on the experimental results on which they are built. Dynein motion is characterized by its ability to move helically [17] as well as backwards [15]. However, in terms of navigating obstacles it is the latter process that is most important. Our simulations reveal that changes in off-axis stepping probability have little effect on the statistics of motor protein motion (Fig 6), but the introduction of even a small probability of backwards

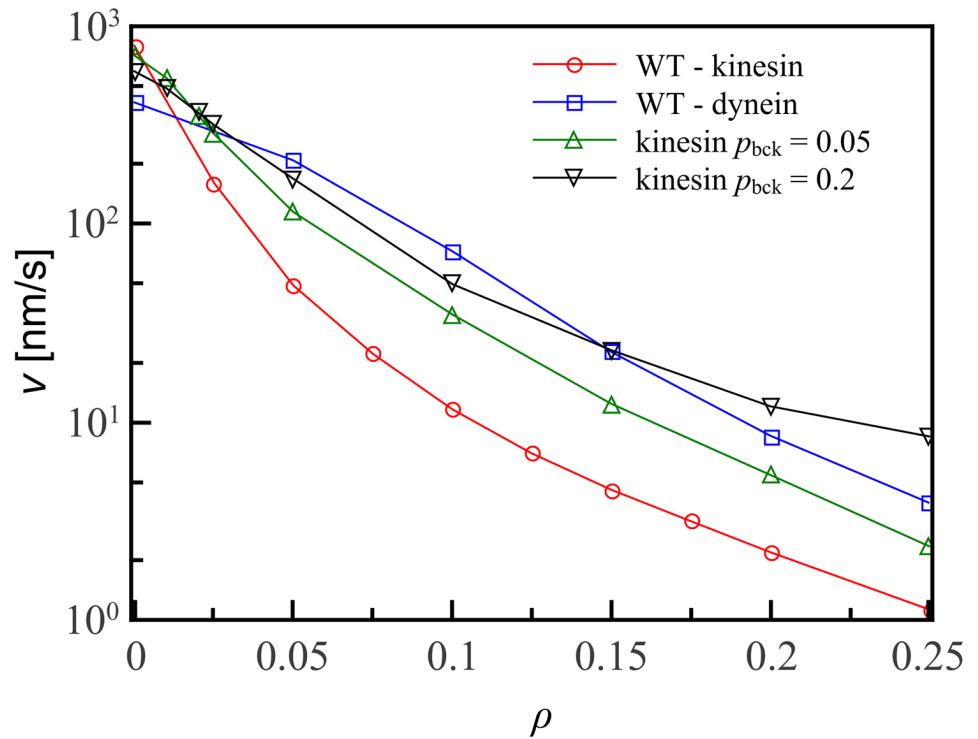

**Fig 7. When kinesin is allowed to take backwards steps its obstacle navigational abilities increase drastically.** The effect is especially dramatic for large decoration fraction $\rho$. When kinesin can take backwards steps with the same probability $p_{bck} = 0.2$ as wild-type dynein, for $\rho > 0.15$ the backwards-stepping kinesin is faster than wild-type dynein. Lines are guides to the eye; errorbars are smaller than the symbol size.

doi:10.1371/journal.pone.0136945.g007

motion to otherwise regular stepping increases the velocity of the motor protein by an order of magnitude for large decoration fractions $\rho \approx 0.25$ (Fig 7). It would be interesting to compare our numerical results on the dynamics of motors on decorated tracks with experiments in vivo. Unfortunately, no direct measurements of obstacle densities in vivo are available to the best of our knowledge.

An interesting future direction will be to extend this model to capture more complex collective effects. Cargo is typically transported by multiple motors which may collaborate or compete (see ref. [33] for a recent review). Indeed, collective motion of motor proteins in crowded environments displays interesting behaviors such as burstiness and broad distributions of run lengths [34, 35]. How these behaviours are affected by the stepping of individual motors is yet to be elucidated.

## Supporting Information

**S1 Video. Sample trajectory of a dynein motor on a clean track.**
(MOV)

**S2 Video. Sample trajectory of a helical kinesin motor on a clean track.**
(MOV)

**S3 Video. Sample trajectory of a dynein motor on a decorated track.**
(MOV)

**S4 Video. Sample trajectory of a helical kinesin motor on a decorated track.**
(MOV)

## Author Contributions